\documentclass[pra, amsmath, showpacs, superscriptaddress, twocolumn, sort&compress, floatfix, amssymb, noeprint, longbibliography]{revtex4-1}
\usepackage{graphicx,color}
\usepackage{mathptmx, textcomp, float}
\usepackage{braket,amsfonts}
\usepackage[hidelinks]{hyperref}

\begin{document}

\title{Collisionally inhomogeneous Bose-Einstein condensates with a linear interaction gradient}

\author{Andrea Di Carli, Grant Henderson, Stuart Flannigan, Craig D. Colquhoun, Matthew Mitchell, Gian-Luca Oppo, Andrew J. Daley, Stefan Kuhr, Elmar Haller}
\affiliation{Department of Physics and SUPA, University of Strathclyde, Glasgow G4 0NG, United Kingdom }
\date{\today}


\begin{abstract}
We study the evolution of a collisionally inhomogeneous matter wave in a spatial gradient of the interaction strength. Starting with a Bose-Einstein condensate with weak repulsive interactions in quasi-one-dimensional geometry, we monitor the evolution of a matter wave that simultaneously extends into spatial regions with attractive and repulsive interactions. We observe the formation and the decay of soliton-like density peaks, counter-propagating self-interfering wave packets, and the creation of cascades of solitons. The matter-wave dynamics is well reproduced in numerical simulations based on the nonpolynomial Schr\"odinger equation with three-body loss, allowing us to better understand the underlying behaviour based on a wavelet transformation. Our analysis provides new understanding of collapse processes for solitons, and opens interesting connections to other nonlinear instabilities.
\end{abstract}

\maketitle

Collisionally inhomogeneous fluids exhibit spatially varying interactions between their particles. They frequently occur at interfaces, where interaction properties change due to a variation of an external potential or due to a change of the fluid's composition. Examples for fluids at interfaces with a collisional inhomogeneity are liquid-vapour surfaces, and material junctions in condensed-matter physics. Studying quantum gases at boundaries is a current experimental challenge, with the goal to simulate mechanisms that modify transport, such as Andreev-like reflections for a gas of electrons \cite{Safi1995,Daley2008}.

In this Letter, we provide a first experimental study of the dynamical properties of a quantum gas with a \textit{linear} collisional inhomogeneity. Specifically, we explore Bose-Einstein condensates (BEC) with spatially mixed interactions, i.e., wave packets that expand simultaneously into regions with attractive and repulsive interaction. Starting from a trapped BEC in a region with weak repulsive interaction, we monitor its spreading along a vertical guiding potential into a region with attractive interactions. We study the density profile of the wave packet as it evolves and expands, and we observe the formation and decay of a soliton-like peak, counter-propagating self-interfering wave packets, and cascades of solitons. Modelling quantitatively the underlying nonlinear dynamics numerically, we use wavelet transformations to extract position and momentum information. This allows us to understand the collapse dynamics, and identify the vital role three-body processes play in controlling these processes in experiments. This opens intriguing potential connections to related phenomena, including a recent experimental study \cite{Khamehchi2017} showing shock waves and soliton trains as a result of an effective negative mass in spin-orbit-coupled BECs, in which the existence of counter-propagating wave packets was also predicted \cite{Colas2018}.

Several methods were previously proposed to create quantum gases with position-dependent interactions. Optical Feshbach resonances \cite{Fedichev1996c,Fatemi2000a,Theis2004}, and optically controlled magnetic Feshbach resonances \cite{Bauer2009,Clark2015}, can be used to tune the interaction strength with position-dependent properties of laser beams \cite{Yamazaki2010a,Yan2013a,Clark2015,Arunkumar2019}. However, optical control methods are typically aimed to change the interaction strength on short length scales, and they often suffer from loss and heating of the gas. Our scheme is based on magnetic Feshbach resonances and magnetic field gradients as proposed in Refs.~\cite{Theocharis2005a,Theocharis2006,Niarchou2007a,Rodrigues2008}. It allows for a long observation time of hundreds of milliseconds, which is necessary to study the impact of small, spatial changes of the interaction strength on the dynamical evolution of gas. We apply a magnetic field gradient $\partial_z B$ along the $z$-direction, $B(z) = B_\text{off} + \partial_z B\cdot z$, which directly maps the field-dependent s-wave scattering length, $a(B)$, to position space. A drawback of this scheme is the creation of a strong, constant force due to the field gradient and the resulting Zeeman shift of the atomic energy levels. In our experimental setup, we apply a vertical magnetic field gradient with a value that compensates the gravitational acceleration, thus effectively cancelling both forces.

\begin{figure}[t]
\centering
  \includegraphics[width=0.49\textwidth]{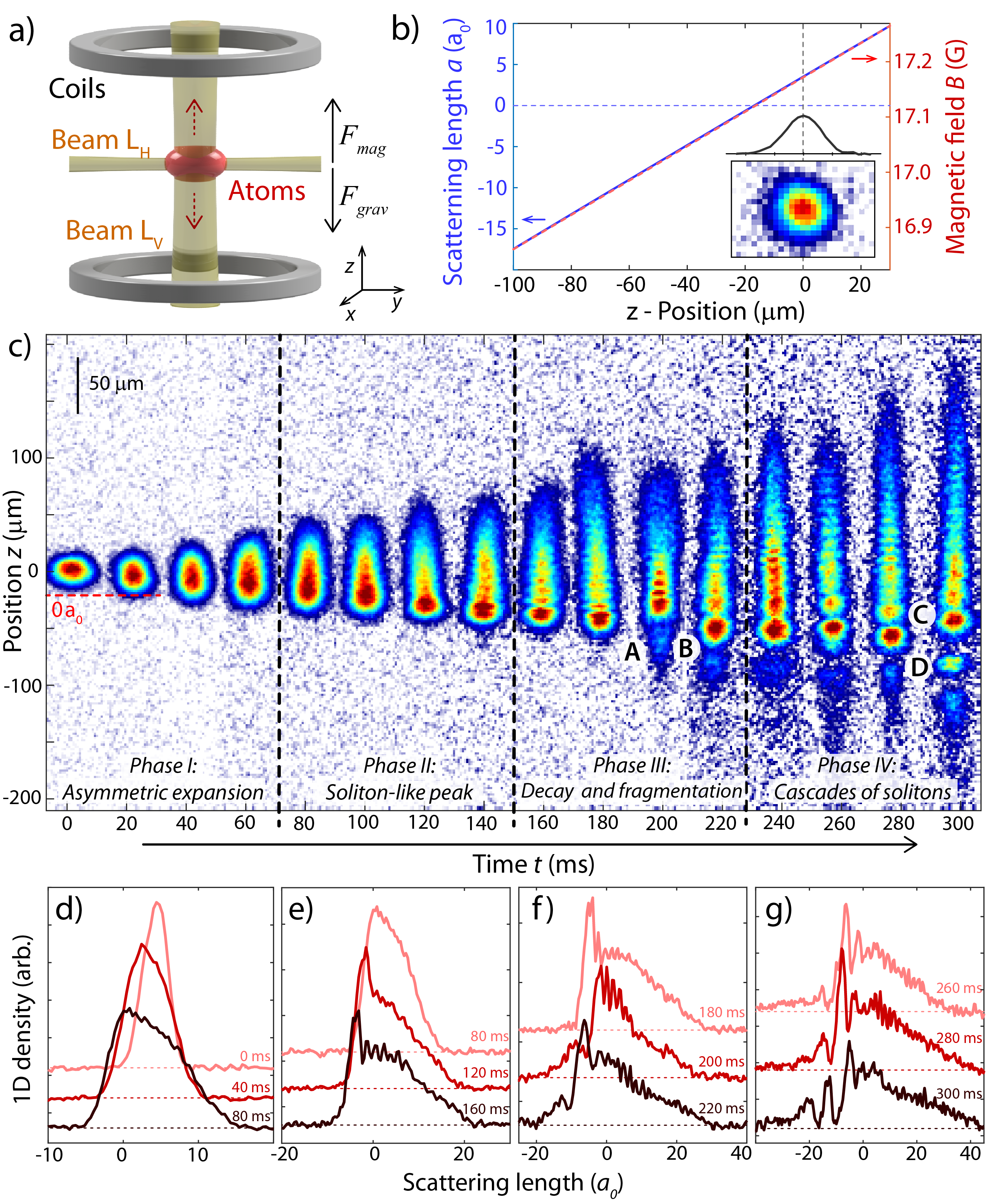}
  \vspace{-2ex}
 \caption{(a) Experimental setup with coils and with the vertical and horizontal laser beams $L_V$ and $L_H$. The gravitational force, $F_\text{grav}$, is cancelled by the magnetic force $F_\text{mag}$. (b) Position-dependence of the scattering length (left axis) and the magnetic field (right axis). Inset: absorption image of the initial density distribution at position $z=0\,\mu$m after a free expansion time of 25\,ms. (c) Absorption images of the density distribution after a variable evolution time $t$ and an expansion time of 25\,ms. The colormap of each image is re-scaled using the peak-density to enhance the visibility of the evolving structures in the density profiles. (d-g) Horizontally integrated density profiles of the images in (c). \label{fig:setup}}
\end{figure}

Depending on the structure of magnetic Feshbach resonances, the resulting position-dependent scattering length, $a(z)$, can be a complicated function showing spatial regions with attractive, repulsive and diverging interaction. While $\partial_z B$ is determined by our levitation scheme, we can use $B_\text{off}$ to control the spatial variation of $a(z)$. For a first experimental study, we chose a regime with a linear variation of the scattering length, $a(z) = a_\text{off} + \partial_z a\cdot z$, with a constant spatial gradient $\partial_z a$, and an offset value $a_\text{off}$. The gradient introduces a new length scale to the system, $L_a = |a_\text{off}/\partial_z a|$, which corresponds to a displacement that changes the scattering length by $a_\text{off}$. This length scale $L_a$ can be compared to typical length scales of the matter-wave packet, $L_w$, e.g., to the healing length or to the Thomas-Fermi radius. For $L_a\gg L_w$, the relative change of $a$ is small over the extend of the matter wave, and a propagating wave packet can adapt its size and peak density while preserving its overall shape. For this regime, an analytical solution for a bright soliton was derived based on the perturbed nonlinear Schr\"odinger equation \cite{Theocharis2005a,Theocharis2006}. For a strong interaction gradient with $L_a\le L_w$, the shape of the matter wave is changed and local self-amplifying feedback effects can be expected.

Details of our experimental apparatus and cooling sequence can be found in earlier publications \cite{DiCarli2019, DiCarli2019c}. A BEC of approximately 12,000 cesium atoms \cite{DiCarli2019c} in the Zeeman sub-state $\ket{F=3,m_F=3}$ is trapped in the dipole potential of two crossed laser beams, $L_H$ and $L_V$, with trap frequencies $\omega_{x,y,z}=2\pi\times (40(1),40(1),5.3(1))\,$Hz (Fig.\,\ref{fig:setup}(a)). Two pairs of vertical coils create a homogeneous magnetic field $B_\text{off}$ and a magnetic field gradient $\partial_z B_\text{lev}=31.1\,$G/cm to compensate the gravitational acceleration. The scattering length $a_\text{off}$ at the initial position of the BEC is $+4\,a_0$ with an approximately linear variation $\partial_z a = 0.21\,a_0/\mu$m (Fig.\,\ref{fig:setup}(b)), where $a_0$ is Bohr's radius. We initialize the expansion of the BEC along the guiding laser beam $L_V$ by switching off the beam $L_H$ within $4\,$ms.

Figure\,\ref{fig:setup}(c) shows absorption images of the expanding BEC over an evolution time $t$ of $300$\,ms with the corresponding horizontally integrated one-dimensional density profiles in Figs.\,\ref{fig:setup}(d-g). To facilitate the discussion, we divide the evolution of the spreading wave packet into four phases (I-IV). Phase I is dominated by an asymmetric, interaction-driven expansion of the BEC (Fig.\,\ref{fig:setup}(d)). In Phase II, a sharp matter-wave front develops in the density profile on the attractive side forming a sharp soliton-like peak (Fig.\,\ref{fig:setup}(e)). Periodic ripples in the density profile start to propagate from the attractive side of the wave packet to the repulsive side in Phase III, and the sharp peak in the density profile on the attractive side decays with only a small cloud of remaining atoms (Symbol A at $200\,$ms in Fig.\,\ref{fig:setup}(c)). The residual atoms continue to propagate downwards and a second (Symbol B at $220\,$ms) and a third (Symbol C at $300\,$ms) soliton-like peak are visible. The process of formation and decay of density peaks continues in Phase IV (Fig.\,\ref{fig:setup}(g)), e.g.~with the decay of the second peak (Symbol D at $300\,$ms). We discuss these four phases successively in the following paragraphs.

\begin{figure}[t]
\centering
  \includegraphics[width=0.5\textwidth]{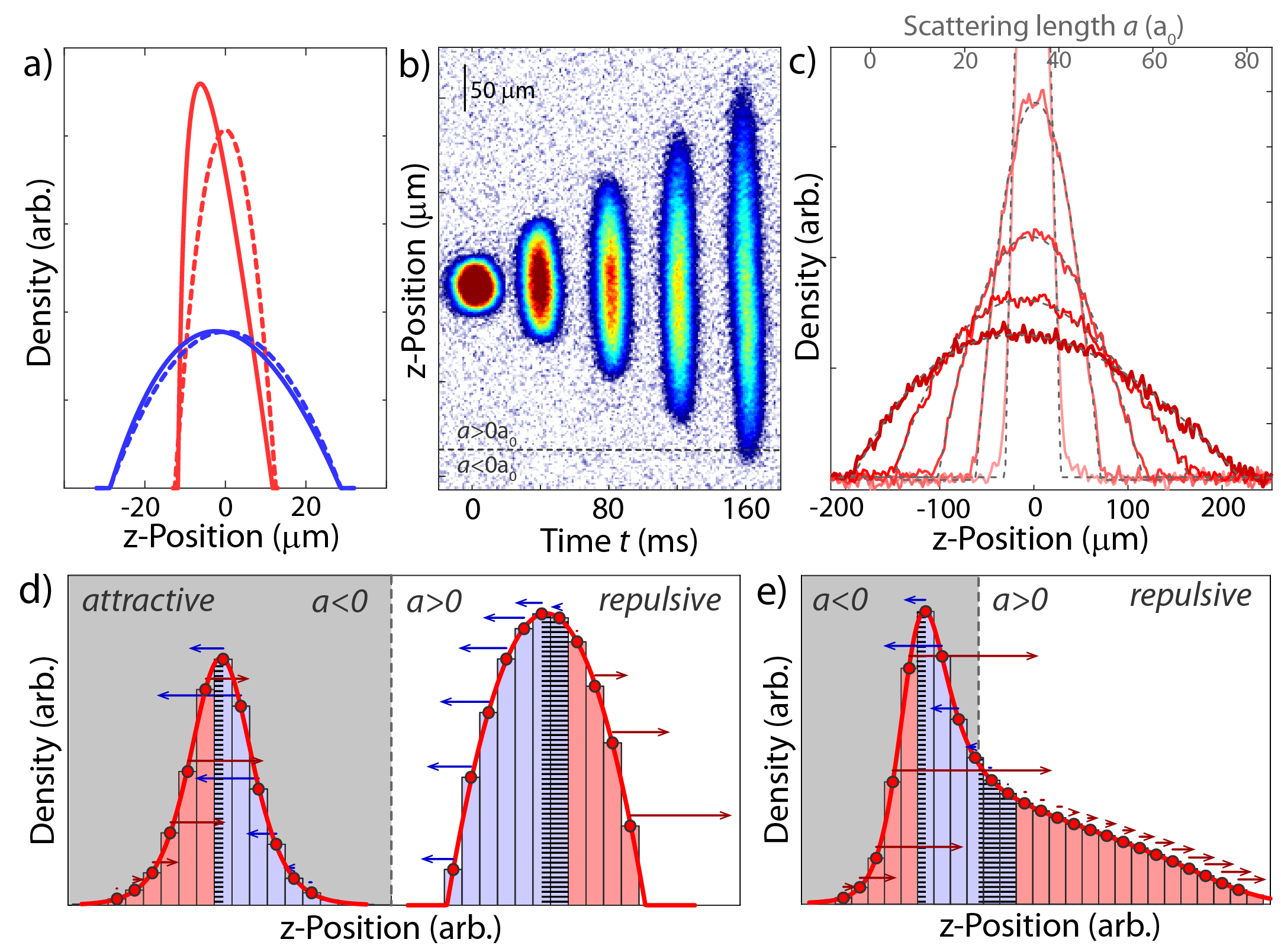}
  \vspace{-2ex}
 \caption{(a) Calculated density profile of the ground state for $N=13,000$, $\omega_r = 2\pi\times 40\,$Hz, $\omega_z = 2\pi\times 5.2\,$Hz, $a=3\,a_0$ (red) and $a_\text{off}=36\,a_0$ (blue) with $\partial_z a=0.21\,a_0/\mu$m (solid lines) and $\partial_z a=0\,a_0/\mu$m (dashed lines). (b) Absorption images and (c) 1D-density profiles of a BEC for varying hold times $t$ and a constant free expansion time of 20\,ms. $a_\text{off}$ is $36\,a_0$ at the initial position of BEC. (d) Illustration of forces on sections of a BEC and soliton, and (e) during the expansion into regions with attractive interaction. Red (blue) patches and arrows indicate sections of the matter wave that are pushed towards smaller (larger) scattering length. \label{fig:expansion}}
\end{figure}

{\bf Phase I. Asymmetric expansion:} Our scattering length gradient causes a linear coupling constant $g(z) =  g_\text{off} + \partial_z g\cdot z$ with an offset $g_\text{off} = 2\hbar \omega_r a_\text{off}$ and a gradient $\partial_z g = 2\hbar \omega_r \partial_z a$. Here, $\omega_r$ is the radial trap frequency of $L_V$ in the horizontal plane. For small gradients with $L_a\gg L_w$ and repulsive interactions, the density profile of the ground state in the Thomas-Fermi approximation is given by
\begin{align}
    n(z) = \max(0, \widetilde{n}(z)), \quad \widetilde{n}(z) = \frac{\mu - V(z)}{g_\text{off}} \left( 1 + \frac{\partial_z a}{a_\text{off}} z\right)^{-1} \hspace{-2ex}, \label{eq:slant}
\end{align}
with an external potential $V(z)$ and a chemical potential $\mu$. The interaction gradient adds a position-dependent scaling factor to the density profile, which, as a result, shows a shift of the peak position and a slant towards the side with smaller scattering lengths (Fig.\,\ref{fig:expansion}(a), red lines).

The slanted density profile of the ground state results in an asymmetric expansion when we remove the vertical confinement. However, a slant can also develop in the expansion profile for a symmetric ground state wave function. We observe this effect in an expansion measurement for repulsive interaction with a larger scattering length ($a_\text{off}=36\,a_0$, Fig.\,\ref{fig:expansion}(b)) and an almost symmetric initial wave function of the ground state (Fig.\,\ref{fig:expansion}(a), blue line). The expansion is driven by the conversion of interaction energy to kinetic energy \cite{Kraemer2004}, with an asymmetry due to the position-dependent interaction. We find excellent agreement of the density profiles (Fig.\,\ref{fig:expansion}(c), red lines) with Eq.\,\ref{eq:slant} (Fig.\,\ref{fig:expansion}(c), gray dashed lines) assuming a shape preserving expansion \cite{SuppMat}.

{\bf Phase II. Soliton-like peak:}
In addition to the asymmetric expansion, a further deceleration of the atoms occurs when the wave packet crosses the point of zero interaction (80\,ms, Fig.\,\ref{fig:setup}(e)) and forms a sharp matter-wave front, similar to a single soliton. Here, we provide an intuitive explanation for the coherent formation of the single soliton-like peak.
A full numerical simulation of the time evolution is presented when discussing Phase III.

When neglecting the kinetic energy in the 1D-GPE, the force $F$ on a section of the wave packet at position $z$ is proportional to the spatial derivative of the interaction energy $n(z)g(z)$ with
\begin{align}
    F \sim - \partial_z n(z)\cdot g(z) - n(z)\cdot \partial_z g. \label{eq:forces}
\end{align}
The first term in Eq.\,\ref{eq:forces}, depending on the density gradient $\partial_z n(z)$, causes the spreading of the wave packet for repulsive interaction and a contraction for attractive interaction. The second term, which depends on our interaction gradient, always accelerates the wave packet towards smaller scattering lengths. As a result, the forces are not spatially symmetric, but push the section in the middle of each wave packet towards smaller scattering lengths (Fig.\,\ref{fig:expansion}(d), hatched areas). Patch colors and arrows indicate the forces acting on different sections of the wave packet in Fig.\,\ref{fig:expansion}(d) as prescribed by Eq.\,(\ref{eq:forces}). A soliton-like peak forms when a wave packet spreads into the region with attractive interaction (Fig.\,\ref{fig:expansion}(e)), because forces on sections with an increasing density gradient change sign and counteract the expansion, while the second term in Eq.\,(\ref{eq:forces}) continues to push atoms towards smaller scattering lengths (Fig.\,\ref{fig:expansion}(e), hatched areas).

Bright matter-wave solitons have been created by seeding modulational instabilities in a dense background gas with noise or interferences \cite{Carr2004,Carr2004a,Nguyen2017}, and by shaping the density profile with an external potential before an interaction quench \cite{Khaykovich2002a,Marchant2013,Gosar2019,DiCarli2019c}. Here, in contrast, the single soliton-like peak emerges due to the propagation of the matter wave in the linear interaction gradient and without the need of additional seedings or quenches.

\begin{figure}[t]
\centering
  \includegraphics[width=0.5\textwidth]{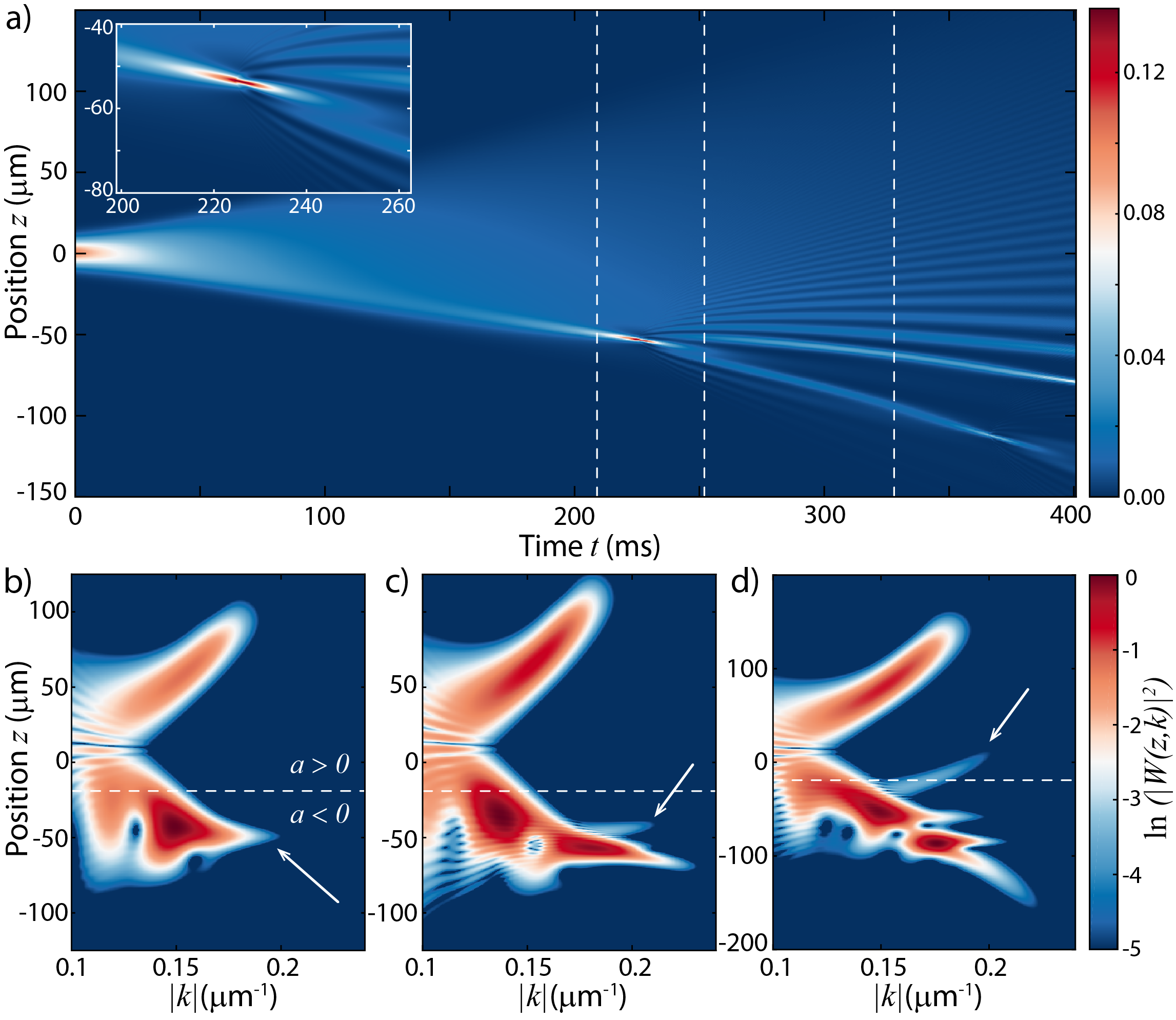}
  \vspace{-2ex}
 \caption{(a) Numerical simulation of the evolution of the density profile using NPSE with three-body loss (1D-density in atoms/$\mu$m). Inset: Zoomed region of the growing and then decaying soliton. (b) Wavelet decomposition \cite{SuppMat} of the atomic wave function at $t=209$\,ms (first vertical line in (a)), showing the emerging soliton (white arrow). (c) Wavelet decomposition at $t=252$\,ms (second vertical line in (a)), after the appearance of the density ripples. The arrow indicates the counter-propagating wave packet. (d) Wavelet decomposition at $t=328$\,ms (third vertical line in (a)), after the counter-propagating wave packet identiﬁed by the arrow penetrates the region of repulsive interactions ($a>0$). \label{fig:simulation}}
\end{figure}

{\bf Phase III. Decay and fragmentation of the soliton-like peak:}
The soliton-like peak propagates towards regions with stronger attractive interaction, while growing in height and shrinking in size as predicted in Ref.\,\cite{Theocharis2005a}. At approximately $t=140\,$ms (Fig.\,\ref{fig:setup}(c)), periodic ripples appear in the density distribution and extend, with a counter-propagating flow, into the region of repulsive interaction. The soliton-like peak decays at approximately $t=200$\,ms with a small cloud of atoms appearing below its former position. We employ numerical calculations based on the nonpolynomial Schr\"odinger equation (NPSE) \cite{Salasnich2002,SuppMat} to simulate the complete evolution of the matter waves (Fig.\,\ref{fig:simulation}). The NPSE alone with a linear dependence of the scattering length reproduces the expansion and soliton-like peak formation well in phases I and II. When the local density becomes large, it is necessary to include a three-body loss term in the model \cite{Santos2002d, Kagan1998, Saito2001,Everitt2017a}.

In the simulation, the  peak density of the propagating soliton increases due to a shrinking of its width (Fig.\,\ref{fig:simulation}(a)), until the three-body loss term increases and prevents a rapid collapse of the soliton \cite{Saito2001,Everitt2017a}. To analyse the effect of the soliton's subsequent decay, we employ a Gabor wavelet decomposition of the wave function \cite{SuppMat} similar to what was used in Ref.\,\cite{Colas2018}. In contrast to a Fourier spectrum which uses delocalized sine and cosine functions, the wavelet decomposition provides information about the location, direction and velocity of motion of wave packets at a given time. In the wavelet decomposition at $t=209$\,ms the soliton is visible as a peak in the region of attractive interactions (arrow in Fig.\,\ref{fig:simulation}(b)). After the soliton starts to decay at $t=220\,$ms, a counter-propagating wave packet develops on the side of the soliton (arrow in Fig.\,\ref{fig:simulation}(c)) at $t=252\,$ms) and then penetrates the region of repulsive interactions (arrow in Fig.\,\ref{fig:simulation}(d) at $t=328$ms). We interpret this wave packet
as a partial reﬂection of the incoming atoms on the sharp edges of the now decaying soliton. The counter-propagating wave packet coherently interferes with the atoms moving towards negative scattering lengths, giving rise to the ripple pattern in the atomic density observed in both the attractive and repulsive interaction regions. We observe no change of the ripple pattern at the zero-crossing of the scattering length, where the interaction-dependent speed of sound in the gas vanishes (dashed lines in Fig.\,\ref{fig:simulation}(b-d)).

\begin{figure}[t]
\centering
  \includegraphics[width=0.49\textwidth]{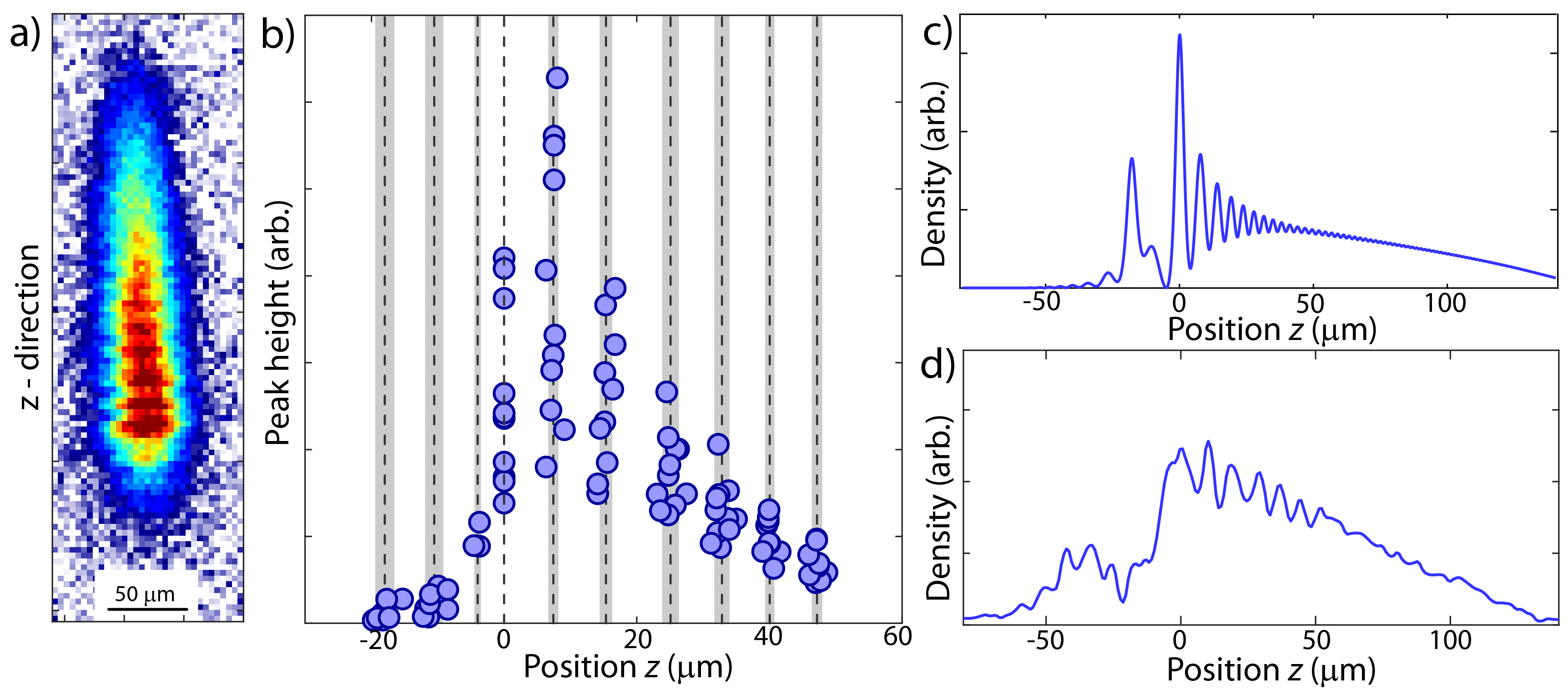}
  \vspace{-3ex}
 \caption{Ripple pattern in the density profile for approximately $22,000$ atoms. (a) Absorption image of density profile after a free expansion time of 25\,ms. (b) Positions of density peaks for multiple absorption images. Mean position and standard deviation of the numbered peaks are indicated by dashed lines and by gray areas, respectively. (c) Simulated density distribution at $\Delta T=40$\,ms after the decay of the soliton. (d) Measured density distribution at $\Delta T=40$\,ms \cite{SuppMat}.\label{fig:ripples}}
\end{figure}

In Fig.\,\ref{fig:ripples}, we study the ripples of the density profile in the experiment and in the simulation. We find qualitative agreement between the density profiles of the simulation and the experiment (Fig.\,\ref{fig:ripples}(c,d)). In the experiment, the pattern shows fluctuations of the peak density and of the positions of the ripples, but the distances between the ripples are maintained from shot-to-shot. To test this, we identify for each absorption image the peak positions relative to the position of the first large density peak next to the decaying soliton (Fig.\,\ref{fig:ripples}(b)). The distances between the peaks are close to $8\,\mu$m, with a slow increase in time (see details in \cite{SuppMat}).

{\bf Phase IV. Cascades of soliton-like peaks:}
In the simulation, we observe that ripples in the region of repulsive interaction do not grow with propagation time and their amplitudes decay the further away they are from the point of zero scattering length. This is markedly different to atoms in regions of attractive interaction, where experiment and simulations show the formation of additional soliton-like peaks for long hold times. The peak-formation process of Phase II repeats itself as atoms are continuously pushed towards the region with attractive interaction, and, for large initial densities, the density ripples of Phase III attract the surrounding medium, shrink in size, grow in height, and form new separate soliton-like peaks (Fig.\,\ref{fig:simulation}(a)). This is in agreement with the appearance of a second and a third soliton-like peak in the experimental data during Phases III and IV (Symbols B and D, Fig.\,\ref{fig:setup}c). We also expect secondary decays to occur, as those soliton-like peaks propagate towards stronger attractive interaction, and new ripple patterns form (Fig.\,\ref{fig:simulation}(a)). A related scheme to create multiple solitons was proposed in Ref.\,\cite{Rodas-Verde2005}, based on a careful tuning of position-dependent interactions and an optical dipole trap. Our results show that even without the dipole trap, cascades of solitons form due to the interaction gradient.

In conclusion, we provided a first experimental study of the dynamical evolution of collisionally inhomogeneous matter waves with a linear interaction gradient. We observed the time evolution of a BEC in quasi-1D geometry, as it expands with repulsive iteration into non-interacting and attractive regions. Four characteristic phases were identified in the time evolution, i.e.~an asymmetric expansion, the coherent formation and decay of a soliton-like peak, the generation and propagation of counter-propagating coherent waves, and the creation of cascades of solitons. Utilizing our spectrum of magnetic Feshbach resonances, we expect that our experimental method can be extended beyond a linear interaction gradient and small scattering lengths, to study matter wave transport in complex potentials with position-dependent loss and rapid interaction changes as they occur at boundaries and interfaces.

\vspace{1ex}

We acknowledge support by the EU through ``QuProCS'' (GA 641277) and the ETN ``ColOpt" (GA 721465), and by the EPSRC Programme Grant ``DesOEQ" (EP/P009565/1).  AdC acknowledges financial support by EPSRC and SFC via the IMPP. GH acknowledges financial support by The Carnegie Trust via the Vacation Scholarship scheme.

\bibliography{SolGradBib,SuppMatTex}

\onecolumngrid\newpage\twocolumngrid

\section*{Supplemental Material}

\section{Numerical simulations}

\subsection{Non-Polynomial Schr\"odinger Equation}
For sufficiently low particle density $n$ and s-wave scattering length $a$, with $n |a|^3 \ll 1$, we can describe the dynamics of a Bose-Einstein condensate (BEC) within a mean field approximation using the 3D Gross-Pitaevski equation (GPE) \cite{PhysRevLett.88.210403},
\begin{equation}\label{GPE_Diff}
 \begin{split}
i\hbar\frac{\partial}{\partial t}\psi(\mathbf{r},t)=& \left[-\frac{\hbar^{2}}{2m}\nabla^{2}+ V(\mathbf{r})\right.\\
&\left. +g(z)N\left|\psi(\mathbf{r},t)\right|^{2}\right]\psi(\mathbf{r},t).
 \end{split}
\end{equation}
Here, the atomic wave function $\psi(\mathbf{r},t)$ describes the macroscopically populated state, and $g(z)$ is the position-dependent coupling constant. For our experimental gradients of $g(z)$, the change of two-body scattering parameters is negligible over the effective range of interparticle interactions, and we can use a spatially varying scattering length $a(z)$, with $g(z) = 4\pi \hbar^2 a(z)/ m$. The external potential $V(\mathbf{r})$ in Eq.\,(\ref{GPE_Diff}) describes a three-dimensional harmonic trap with anisotropic trap frequencies.

For a cylindrical trap geometry, an effective 1D equation for the dynamics of the BEC can be derived by expressing the wave function in terms of the single particle eigenstates of the radial harmonic oscillator. By using a standard Gaussian ansatz, the wave function $\psi(\mathbf{r},t)$ consists of a Gaussian function in the radial direction and an arbitrary component $f(z,t)$, in the longitudinal direction,
\begin{equation}\label{ansatz}
\begin{split}
\psi(\mathbf{r},t) &=  f(z,t) \phi(x,y,\sigma) \\
&= f(z,t) \frac{1}{\sqrt{\pi}a_{r}\sigma(z,t)}\exp\left[-\frac{(x^{2}+y^{2})}{2a_{r}^{2}\sigma(z,t)^{2}}\right],
\end{split}
\end{equation}
where $a_r$ is the harmonic oscillator length in the radial direction and $\sigma(z,t)$ is the spatially dependent width.

Assuming that $\nabla^2_z\phi(x,y,\sigma) \ll  \left( \nabla^2_x + \nabla^2_y \right) \phi(x,y,\sigma)$, we can use the Euler-Lagrange equations to derive the equation of motion. Following \cite{PhysRevA.65.043614}, we obtain the non-polynomial Schr\"odinger equation (NPSE) given by
\begin{equation}\label{PDE}
 \begin{split}
i\hbar \frac{\partial}{\partial t}f(z,t) &= \left[ -\frac{\hbar^{2}}{2m}\frac{\partial^{2}}{\partial z^{2}}+V(z) \right. \\
&+\frac{g(z)N}{2\pi a_{r}^{2}\sigma(z,t)^{2}}\left|f(z,t)\right|^{2} \\
&\left. +\frac{\hbar\omega_{r}}{2}\left(\sigma(z,t)^{2}+\frac{1}{\sigma(z,t)^{2}}\right) \right] f(z,t),\\
 \end{split}
\end{equation}
with $\sigma(z,t)$ defined by
\begin{equation}
\sigma(z,t)^2 = \sqrt{1 + 2a(z)N|f(z,t)|^2},
\end{equation}
where $N$ is the total number of atoms and $\omega_r={\hbar/(m a_r^2)}$.

In the case of a constant radial width with $\sigma(z,t)^2 = 1$, Eq.\,(\ref{PDE}) corresponds to the 1D GPE. We use the NPSE in this article instead of the more common 1D GPE, because it provides a more accurate approximation of the shape and phase of the wave packet of the full 3D GPE (Eq.\,\ref{GPE_Diff}) over a wide range of scattering lengths \cite{PhysRevA.65.043614}.

\subsection{Three-Body Loss}

Following Ref.~\cite{Kagan1998}, we include the effects of particle loss through three-body recombination via a short range process
\begin{equation}
H_{3B} = -i \frac{\hbar L_3}{6} \int d\mathbf{r} |\psi(\mathbf{r},t)|^6,
\end{equation}
where $L_3$ is the three-body loss parameter. In our model, we add this term to the right-hand side of the effective NPSE (Eq.\,\ref{PDE}) as
\begin{equation}
 -i \frac{\hbar L_3 N^2}{6\pi^2a_r^4\sigma(z,t)^4} |f(z,t)|^4.
\end{equation}

Within these approximations we find that we can capture all qualitative features observed in the experiment. However, the optimal condition for $\sigma(z,t)$ to minimise the effective 1D action can have non-negligible corrections for large values of $|a|$ and $L_3$.

\subsection{Wavelet Decomposition}

For a complex wave function $f(z)$, the wavelet transformation provides its decomposition in position $z$ and wave number $k$ \cite{Simonovski2003}. The wavelet decomposition used in our analysis is given by
\begin{equation}
W(z,k) = \sqrt{k} \int_{-\infty}^{+\infty} f(z') G^*[k(z'-z)] \; dz',
\label{eq:WD}
\end{equation}
where $k$ is the scale, $z$ the translation parameter and $G(s)$ is the Gabor/Morlet wavelet
\begin{equation}
  G(s) = \pi^{-1/4} \exp(i\omega s) \exp(-s^2/2),
  \label{eq:WDGM}
\end{equation}
with $\omega$ being the wavelet's frequency. In our simulations, we have applied the wavelet decomposition Eqn.\,(\ref{eq:WD})-(\ref{eq:WDGM}) to the NPSE wavefunction $f(z,t)$ at fixed times $t$ with $\omega$ optimised to generate equal variance in time and frequency (see \cite{Simonovski2003} for definitions). Figures 3(b,c,d) in the main text display  $|W(z,k)|^2$ at the given times, showing different wave packets that move at a given position $z$ with different wave numbers $k$.

\section{Experimental methods}

\subsection{Asymmetric expansion in phase I}

\begin{figure}[h]
\centering
 \includegraphics[width=0.47\textwidth]{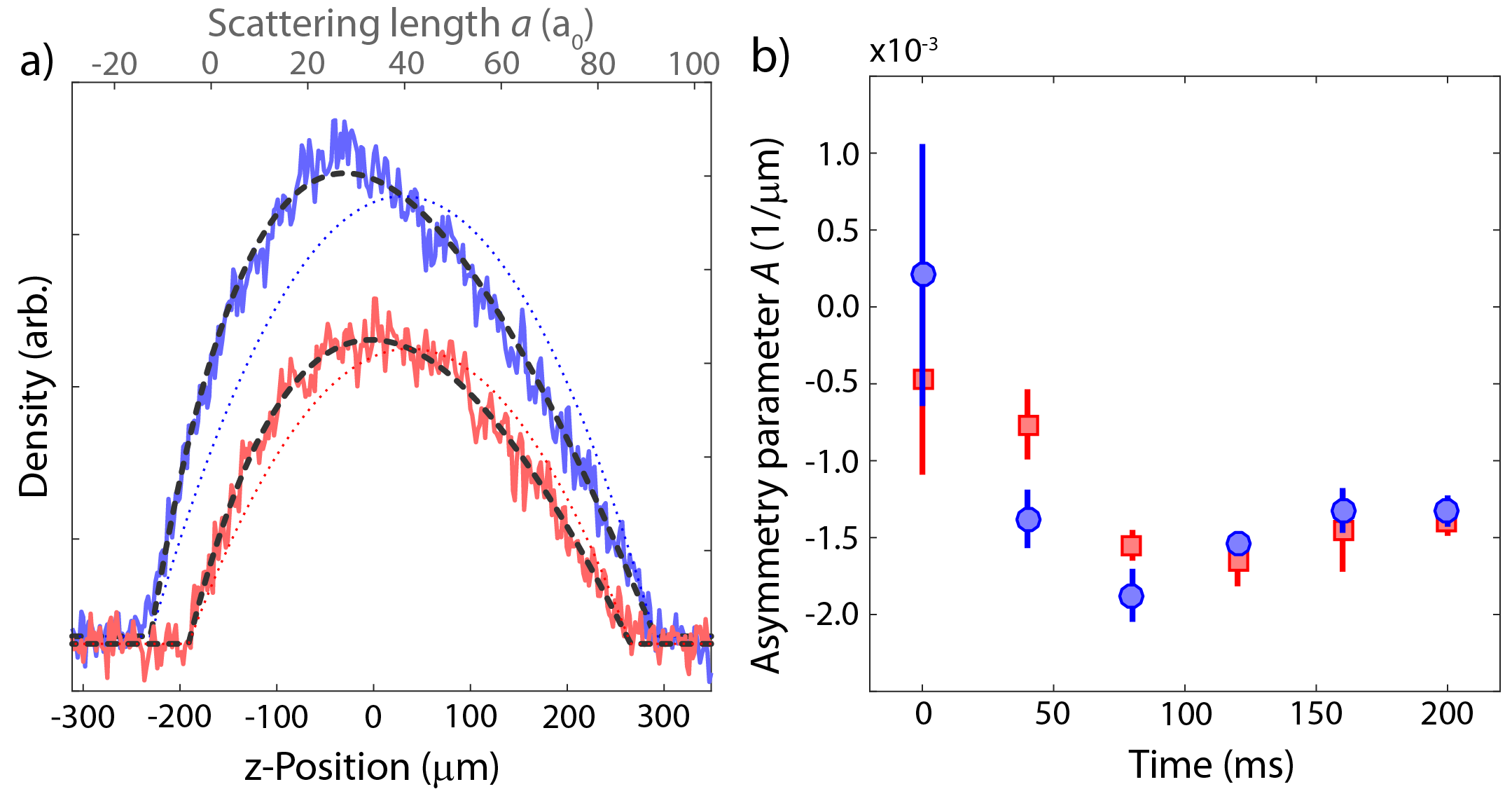}
 \caption{Asymmetric expansion in quasi-1D with a gradient of repulsive interactions. (a) Density profile after an expansion time of 200\,ms in the guiding potential for 33,000 atoms (blue line) and 18,000 atoms (red line). Dashed lines: fit with Eq.\,(\ref{eq:fitting}), dotted lines: same fit result with $A=0$. The scattering length at $z=0\,\mu$m is $a_\text{off}=36\,a_0$. (b) Asymmetry parameter $A$ for increasing expansion time, and approximately 33,000 atoms (blue circles) and 18,000 atoms (red squares). \label{fig:asym}}
\end{figure}

We studied the density profile of a matter wave packet during its expansion in a quasi-1D system with a linear gradient of repulsive interactions. The expansion of the wave packet in the guiding potential of laser beam $L_V$ is initialized by switching off the beam $L_H$, and we measure the density profile after a varying expansion time of 0-200\,ms. For repulsive interactions, $a_\text{off}=36\,a_0$, the expansion is driven by a conversion of interaction energy to kinetic energy. Due to the interaction gradient $\partial_z a$, we expect a position-dependent interaction energy to drive an asymmetric expansion. This asymmetry is clearly demonstrated by the density profiles in Fig.\,\ref{fig:asym}(a) for 33,000 atoms (blue line) and for 18,000 atoms (red line).

Our fit function for the 1D-density distribution is based on Eq.\,(1) in the main text, with
\begin{align}\label{eq:fitting}
 \begin{split}
    n(z) = B \max\left( 0, 1-\frac{(z-D)^2}{C^2} \right) & \frac{1}{1 +  A (z-D)} \\
     & \qquad\qquad + F.
\end{split}
\end{align}
The fit parameters $A,B,C,D,F$ measure the asymmetry, amplitude, width, shift and offset of the density profile, respectively. We find excellent agreement of the fit function with our profiles (dashed lines, Fig.\,\ref{fig:asym}(a)). We use the parameter $A$ to quantify the slant of the density profiles. As a reference, we show the same fit functions with parameter $A$ set to zero (dotted lines, Fig.\,\ref{fig:asym}(a)).

We observe a converges of the fit parameter $A$ for long expansion times to values of approximately $-0.0015\, (1/\mu$m) (Fig.\,\ref{fig:asym}b). This convergence is in agreement with our expectation that the interaction energy of the wave packet decreases for long expansion finally resulting in a ballistic motion of the atoms.

{\bf Absorption imaging:} For our experimental parameters, in-situ absorption images of the atoms are optically dense, and we need to add a short period of untrapped, levitated expansion before taking the images (25\,ms). During this period, the atoms expand mainly along the horizontal direction, which was strongly confined by the guiding beam $L_V$ before release. The magnetic field is not changed during this free expansion period to prevent the creation of additional forces that might change the density profile. Compared to the overall evolution time, this period is short, and we neglect it in our data analysis. Finally, we switch off the magnetic field in 1\,ms, and take absorption images at close to zero magnetic field strength.

\subsection{Uncertainties of experimental parameters} \label{uncert}

Our numerical simulations in Fig.\,3 of the main text show good qualitative agreement with the experimental data. For example, the decay of the soliton-like peak occurs at $t=225$\,ms and $z=-54\,\mu$m in the simulation, which matches well to our observations $t=200\,$ms and $z=-50\,\mu$m (Symbol A in Fig.\,1(c)). However, for observation times after the decay of the soliton-like peak, length and time scales start to deviate. We attribute the deviations to an approximate knowledge of initial conditions and simulation parameters, which will be listed in this section.
\begin{itemize}
    \item {\bf Initial density profile of the BEC:} The initial density profile is calculated using a Thomas-Fermi distribution and our experimental parameters. We determine our trap frequencies by measuring center-of-mass oscillations in the dipole trap with 3\% measurement uncertainty. Our shot-to-shot atom number fluctuations are below 5\%, and we estimate the systematic error of the atoms number measurement to be below 10\%. No thermal fraction is detectable for the initial BEC.
    \item {\bf Three-body loss coefficient $L_3$:} The three-body loss coefficient of cesium atoms in the $\ket{3,3}$ hyperfine state is within the range of $5\times10^{-28}$\,(cm$^6$/s) to $5\times10^{-27}$ (cm$^6$/s) for magnetic fields close to $17$\,G \cite{Kraemer:2006aa}. We performed numerical simulations covering $L_3$ parameters from $5\times10^{-28}$ to $5\times10^{-27}$ (cm$^6$/s), in accordance with experimental limits. For the lowest values of $L_3$ we observe that the soliton decay occurs around 220\,ms, and for the highest ones, we saw that the soliton decays approximately 40\,ms later. We found that for a value of $L_3=5\times10^{-28}$ (cm$^6$/s) the time of the soliton decay best matches the experimentally observed ones.
    \item {\bf Scattering lengths $a_\text{off}$ and $\partial _a a$} The magnetic field of the zero-crossing of the scattering length was experimentally determined to be 17.119(2)\,G in a precision measurement using Bloch oscillations \cite{Gustavsson2008}. Our calculation of $a(B)$ close to the zero-crossing is based on Refs.\,\cite{Kraemer2004,Berninger2013}.
    \item {\bf Magnetic fields and external forces:} Our current regulation for the magnetic field generating coils provides a relative current stability of $10^{-6}$ with a similar stability for $B_\text{off}$ \cite{DiCarli2019}. Vertical forces are generated by gravity, by our magnetic field gradient, and by the longitudinal confinement of the laser beam $L_V$ (200\,$\mu$m waist). We balance our vertical forces by minimizing the center-of-mass motion of a BEC with low density and weak, repulsive interactions. Using our residual vertical drifts, an upper limit for the uncertainty of the vertical force is $\sim10^{-5}g$.
\end{itemize}

\subsection{Density ripples}

\begin{figure*}[t]
\centering
 \includegraphics[width=0.8\textwidth]{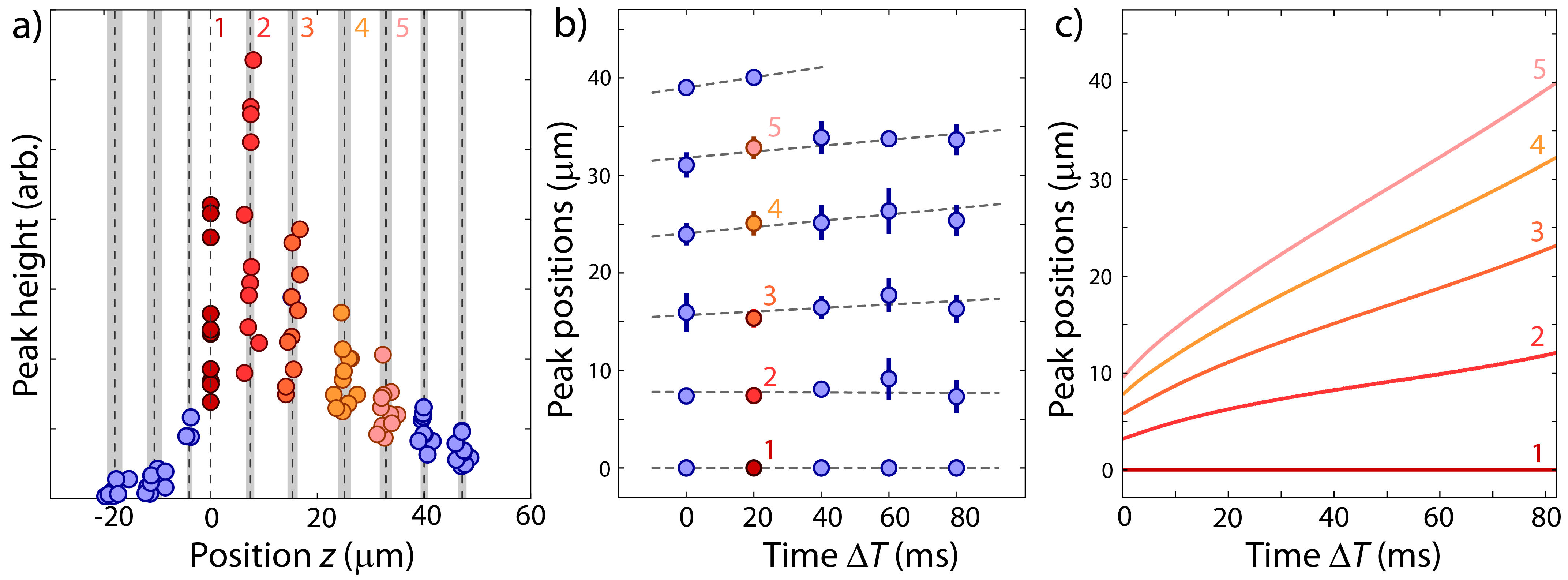}
 \caption[width=1\textwidth]{Ripple pattern in the density profile for approximately $22,000$ atoms. (a) Position of density peaks at $\Delta T=20$\,ms for multiple absorption images. We identify on each image the reference position $z=0\,\mu$m with the position of the first large density peak next to the decaying soliton. Mean position and standard deviation of the numbered peaks are indicated by dashed lines and by gray areas, respectively. (b) Time evolution of the distances between the peaks. Data points for $\Delta T=20\,$ms correspond to peaks with matching colors and numbers in (a). (c) Corresponding distances between the peaks in the numerical simulation. \label{fig:dist_ripples}}
\end{figure*}

In addition to Fig.\,4 in the main text, we provide a comparison of the spacing between the density ripples in the simulation and in the experiment in Fig.\,\ref{fig:dist_ripples}. In the experiment, the pattern shows shot-to-shot fluctuations of the peak density and positions of the ripples, but the distances between the ripples are reproducible. To quantify the average distance between peaks and the reproducibility, we identify the peak positions for each absorption image relative to the position of the first large density peak next to the decaying soliton (Fig.\,\ref{fig:dist_ripples}(a)). Mean positions and standard deviations of the peaks are indicated by dashed lines and by gray areas, respectively. In Fig.\,\ref{fig:dist_ripples}(a), the density ripples with matching positions are numbered (1-5) and indicated by a common color.

Due to the small shot-to-shot fluctuations, the destructive nature of absorption imaging, and our resolution limit, it is difficult to determine the time of decay of the soliton-like peak precisely. Instead, we reference the evolution time $\Delta T$ to a point in time after the decay of the soliton when we can clearly distinguish the ripples (Fig.\,\ref{fig:dist_ripples}b). The positions and standard deviations of the ripples in Fig.\,\ref{fig:dist_ripples}(a) correspond to the data points (1-5) at $\Delta T=20\,$ms. We observe initial distances of approximately $8\,\mu$m between the ripples, with a slow increase in time. In the simulation, the distances between the ripples increase faster, from approximately $3\,\mu$m to $10\,\mu$m in 80\,ms (Fig.\,\ref{fig:dist_ripples}c). We attribute the deviations between experimental data and simulations to different reference times $\Delta T=0$\,ms, and to an approximate knowledge of our simulation parameters (please also see Section \ref{uncert}).

\onecolumngrid
\twocolumngrid

\end{document}